\documentclass[prl, twocolumn]{revtex4}
\usepackage{graphicx} 
\usepackage{dcolumn} 
\usepackage{xcolor}
\usepackage{multirow}
\begin{document}

\title{Emergent Honeycomb Network of Topological Excitations in Correlated Charge Density Wave}

\author{Jae Whan Park$^1$}
\author{Gil Young Cho$^2$}
\author{Jinwon Lee$^{1,2}$}
\author{Han Woong Yeom$^{1,2}$}
\email{yeom@postech.ac.kr}
\affiliation{$^1$Center for Artificial Low Dimensional Electronic Systems,  Institute for Basic Science (IBS),  77 Cheongam-Ro, Pohang 790-7884, Korea.} 
\affiliation{$^2$Department of Physics, Pohang University of Science and Technology, Pohang 790-784, Korea}
\date{\today}

\begin{abstract}

When two periodic potentials compete in materials, one may adopt the other, which straightforwardly generates topological defects. Of particular interest are domain walls in charge-, dipole-, and spin-ordered systems, which govern macroscopic properties and important functionality. However, detailed atomic and electronic structures of domain walls have often been uncertain and the microscopic mechanism of their functionality has been elusive. Here, we clarify the complete atomic and electronic structures of the domain wall network, a honeycomb network connected by Z$_{3}$ vortices, in the nearly commensurate Mott charge-density wave (CDW) phase of 1T-TaS$_{2}$.
Scanning tunneling microscopy resolves characteristic charge orders within domain walls and their vortices. Density functional theory calculations disclose their unique atomic relaxations and the metallic in-gap states confined tightly therein. A generic theory is constructed, which connects this emergent honeycomb network of conducting electrons to the enhanced superconductivity.


\end{abstract}

\maketitle


Competing periodicities and discommensuration phenomena have been widely discussed from one-dimensional (1D) chain models \cite{brau04, fran49, bak82} to surface adsorbate monolayers \cite{pokr79, fain80} and more recently 2D van der Walls stacking of atomic crystals \cite{rock91, push02, chen12}. For example, the commensuration interaction in graphene layers induces various superstructures of discommensuration with exotic electronic property such as unconventional superconductivity \cite{wood14, cao18}. Discommensurations are also popular and important in 2D systems with more complex interactions such as charge ordered or CDW systems, where domain wall networks have been suggested crucial in metal-insulator transitions \cite{wils75, mcmi76}, emerging superconductivity \cite{sipo08, li16, ang12} and the formation of various metastable states for device functionality \cite{stoj14, yosh15, vask15}.

Among diverse 2D materials, exotic low-dimensional electronic properties and domain walls of 1T-TaS$_{2}$ CDW are outstanding \cite{pill00, cler06, cho15} as combined further with the strong electron correlation \cite{faze80} and the strong spin frustration \cite{law17, klan17}.
1T-TaS$_{2}$ is composed of layers of hexagonal tantalum atoms coordinated octahedrally by sulphur atoms. A metallic phase at high temperature transits into an incommensurate CDW phase below 550 K, a nearly commensurate CDW (NC-CDW) below 350 K, and a commensurate CDW (C-CDW) below 180 K \cite{wils75}.
The atomic structure of the C-CDW phase has the unit of the well-known David-star cluster; among 13 Ta atoms of a cluster twelve are distorted toward the central one forming a $\sqrt{13}\times\sqrt{13}$ supercell \cite{brou80}.
One unpaired electron in the central Ta atom falls into the Mott-insulator state by onsite Coulomb repulsion \cite{faze80}.
The NC-CDW phase was reported to consist of C-CDW domains and a honeycomb domain-wall network \cite{wu89, naka77, yama83, wu90, burk91, spij97} with its atomic structure still unknown and its electronic property under debates \cite{sipo08, stoj14, li16}.
Recent experiments reported that various metallic excited states are formed, presumably with highly conducting domain wall networks, by pressure, doping and other control parameters \cite{stoj14, yosh15, tsen15, cho16, ma16, vask16, xu10, Ang12}. These states may be utilized in a ultrafast memory \cite{stoj14} or a memristor device \cite{yosh15}.
The domain walls are also suggested as the origin of superconductivity emerging from those metallic states \cite{sipo08, yu15, liu16, li16}.
However, recent STM experiments observed the nonmetallic nature of domain walls for the voltage-pulse-induced domain-wall phase \cite{cho16, ma16} and isolated domain walls at low temperature \cite{cho17}.
Under these contradictory findings, the characterization of atomic and electronic properties of domain walls in the NC-CDW phase and the general microscopic understanding of the role of the domain wall network for emerging quantum phases and functionalities are highly requested.


\begin{figure*}
\centering{ \includegraphics[width=16.0 cm]{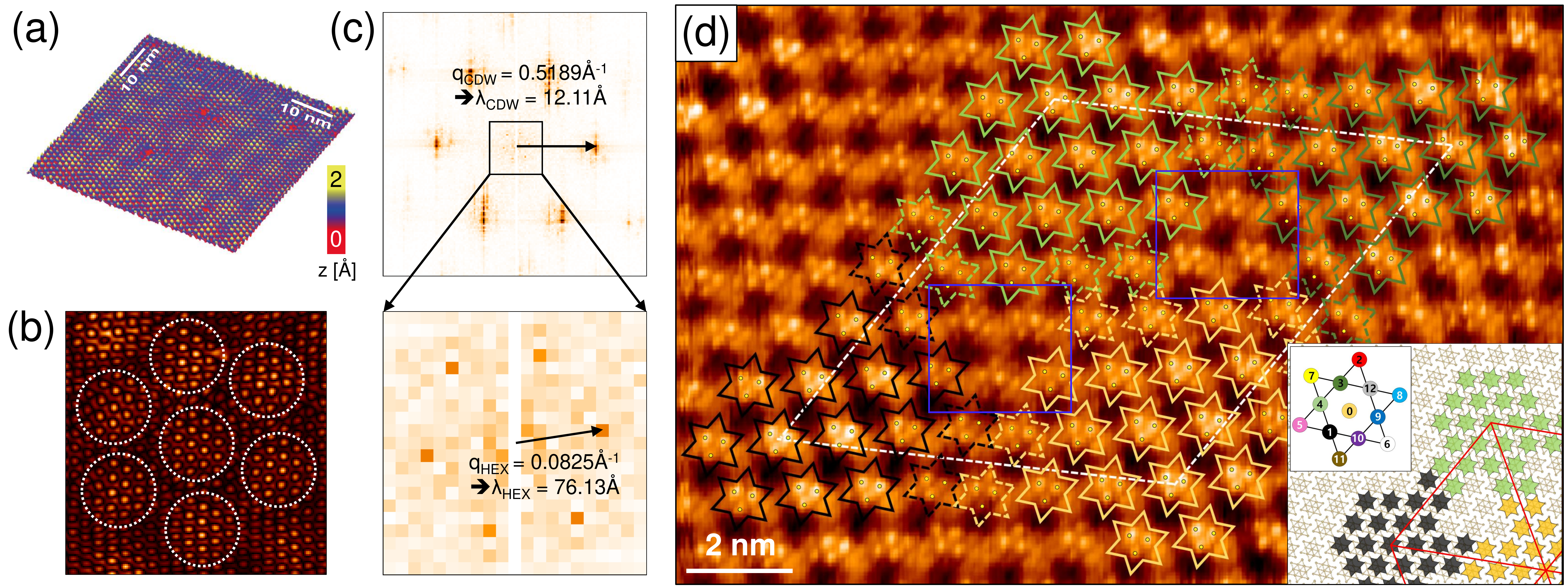} }
\caption{ \label{fig1}
(Color online)
Nearly commensurate charge density wave phase of 1T-TaS$_{2}$ at room temperature.
(a) Constant current STM image (sample bias voltage $V$$_s$ = -10 meV, tunneling current $I$$_t$ = 2 nA, and size $L$$^2$ = 50 $\times$ 50 nm$^2$).
(b) Low-pass filtered image. White circles denote the hexagonal array of domain.
(c) Fast Fourier transform of STM image.
(d) Atomic resolution STM image ($V$$_s$ = -10 meV, $I$$_t$ = 3 nA).
The solid and dashed David star denote the C-CDW clusters in four different domains and domain boundary, respectively.
Filled (yellow) circles denote the top three S atoms in the center of David stars.
Blue squares denote the area of vortex centers.
Inset is a schematic arrangement of David stars for the NC-CDW phase.
}
\end{figure*} 


In this work, we directly resolve atomistic details of discommensuration domain walls of the NC-CDW phase in 1T-TaS$_{2}$ by scanning tunneling microscopy (STM).
By comparison with the experiment, our density functional theory (DFT) calculations identify that the domain-wall consists of particular types of imperfect David-star clusters. 
Domain walls are further connected by Z$_{3}$ topological vortices of domain walls to form a honeycomb network. 
We also identify the atomic structure and the novel charge order within these vortex cores. 
The metallic states of the NC-CDW phase originate from imperfect David stars of domain walls and vortices. A generic theory for the conducting honeycomb network indicates symmertry-protected Dirac bands, quadratic band touchings, and flat bands, which would definitely play important roles in the emerging superconductivity. This work provides unprecedented microscopic understanding of discommensurations in 2D materials with strong interactions, which has wide implication into diverse 2D materials with open possibilities for designing functionalities and quantum phases through the discommensuration engineering.

Figure 1a shows an STM image of the NC-CDW phase in 1T-TaS$_{2}$ at room temperature.
It consists of the domains with relatively high CDW amplitude separated by the low-amplitude regions of domain walls.
A hexagonal array of domains and CDW maxima within each domain are more clearly seen in the low-pass filtered image of Fig. 1(b).
A bright protrusion within a domain corresponds to a single David-star cluster of the C-CDW phase, as revealed more clearly below.
The Fourier transform of the STM image [Fig. 1(c)] exhibits main peaks at $q$$_{CDW}$ = 0.519 {\AA}$^{-1}$  (12.11 {\AA}) in  consistency with the periodicity of C-CDW \cite{cho15}. In addition, the main peaks carry satellites \cite{rits13, hovd16} rotated slightly with respect to the C-CDW vector (see the enlargement) \cite{ishi91, han15}, which have $q$$_{HEX}$ = 0.083 {\AA}$^{-1}$ (76.13 {\AA}) with a rotation of 9.5 $\pm$ 2$^{\circ}$ from the C-CDW peaks. This represents the hexagonal lattice of domains and a honeycomb domain-wall network. 



An atomic resolution STM image reveals the misalignment of David star CDW clusters between domains.
Two adjacent domains with bright protrusions in Fig. 1(d) are separated by the domain-wall region with less bright protrusions. In the domain wall, Daivd-star clusters from two domains, for example, black and yellow ones, are misaligned to share their two outer Ta atoms (see dashed stars). This domain-wall configuration is the first-type domain wall (DW-1) structure with a CDW translation of ${\bf -b}$ (${\bf a}$$\times$${\bf b}$ is the inplane unitcell of the undistorted structure) among the twelve possible configurations due to 13 Ta atoms within one C-CDW unitcell as classified previously \cite{ma16, cho16}. 
This particular configuration is found ubiquitously along all hexagonal directions between domain centers over the whole area probed. 
In the domain area, a single David start exhibits mainly three bright spots arranged in an inverted triangle. 
They correspond to top three S atoms in the center of a David star cluster. 
In contrast, major atomic protrusions in the domain wall are arranged mostly in regular triangles [see the enlarged image in Fig. 2(c)], indicating different types of clusters.
The inset of Fig. 1(d) shows the arrangement of David-star clusters for an ideal hexagonal NC-CDW model as constructed from the STM image.
The distance from one domain center to another is 77.47 {\AA} (22${\bf a}$+2${\bf b}$) with about nineteen perfect David-stars for each domain and with the angles with respect to the C-CDW vector of 9.6$^{\circ}$.
These structural parameters agree well with the present and the previous experiments \cite{wu89}. 
The center position of David-star clusters in each domain is shifted \cite {yama83} and domains are connected with the DW-1 structure, which is the key atomic structure of the NC-CDW phase. Note also that this structure has regions where three domain walls meet [the uncolored David star clusters in Fig. 1(d)]. They are vortices of domain walls as discussed further below.


In order to identify the atomic structure of this characteristic domain wall, we performed DFT calculations.
A simplified supercell of a reduced size from the ideal structure of Fig. 1(d) is used, which consists of five perfect David-star clusters in a domain and a single uniaxial domain wall with the average width of the experiment.
After relaxation, the domain wall is formed with two imperfect David-star clusters, each with twelve Ta atoms, and a link atom between them [Fig. 2(a)].
The simulated STM image of this optimized DW-1 structure is in reasonable agreement with the experimentally observed image of Fig. 2(c) showing three bright protrusions of the inverted triangular shape in the domain region and the regular triangles of less bright protrusions in the domain-wall region.
The energy cost for the formation of a domain wall is as small as 0.03 eV per David star indicating that the phase transition can occur easily from the C-CDW to the NC-CDW phase at room temperature.
The present DW-1 structure differs distinctly from those within the C-CDW phase at low temperature, so called DW-2 or DW-4 structures, which have different displacements between domains and non-metallic density of states \cite {cho17}. These domain-wall configurations at low temperature have a less energy cost by 0.01 and 0.02 eV, which seems consistent with the thermally excited nature of NC-CDW domain walls.


\begin{figure}[t] 
\centering{ \includegraphics[width=8.0 cm]{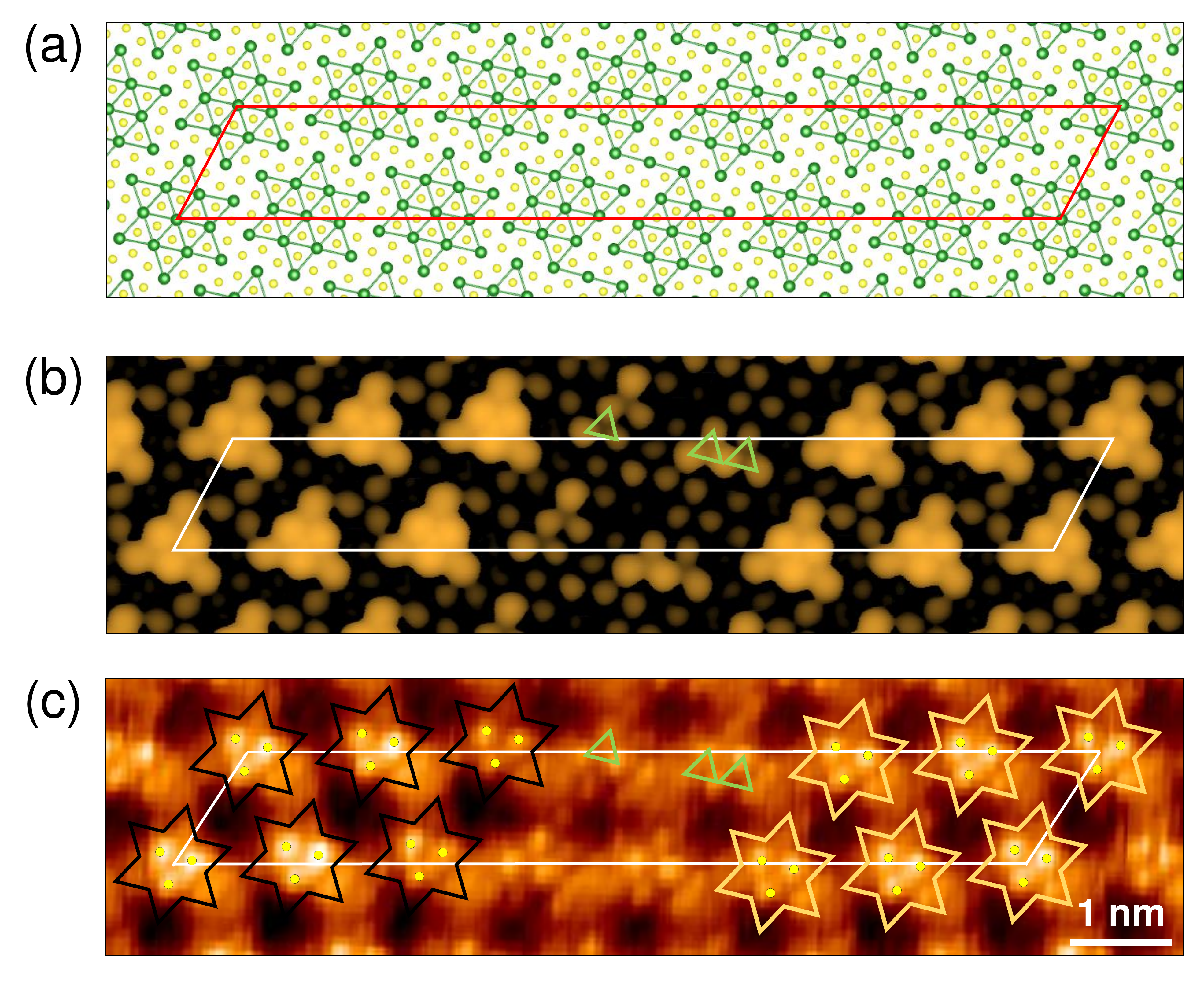} }
\caption{ \label{fig2}
(Color online)
Atomic structure and STM images of DW-1. (a) Atomic structure, (b) Simulated STM image ($V$ = -50 meV,  $\rho$ = 1$\times$10$^{-5}$ $e$/{\AA}$^3$, and (c) Experimental STM image.
The green triangles denote the triangular shape of protrusions in domain-wall area.
}
\end{figure}


\begin{figure}[t] 
\centering{ \includegraphics[width=8.0 cm]{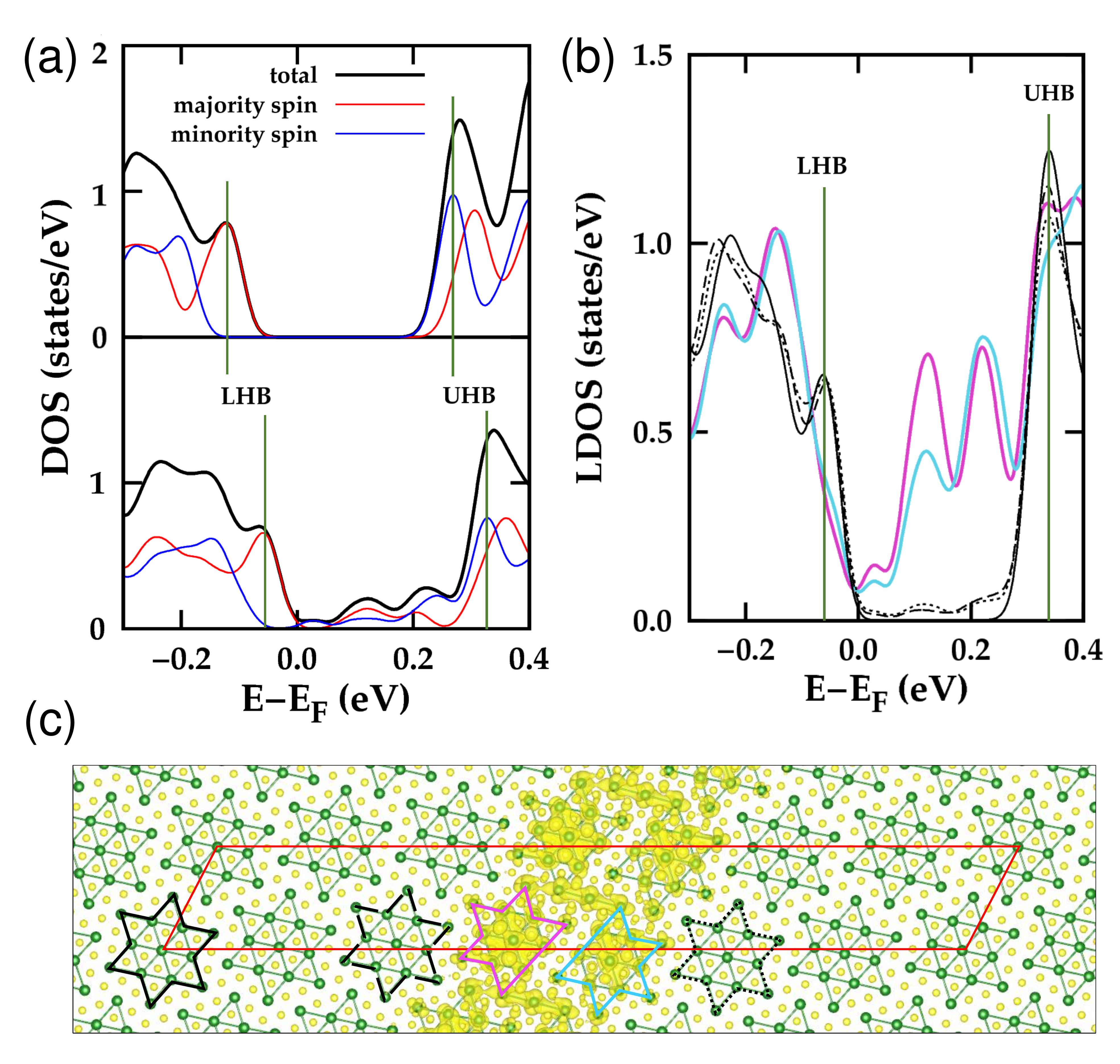} }
\caption{ \label{fig3}
(Color online)
Density of states and charge character of DW-1.
(a) Normalized total DOS of the C-CDW phase (upper panel) as a reference and DW-1 structure (lower panel).
Vertical green lines denote the upper and the lower Hubbard band (UHB and LHB).
(b) Local DOS at all Ta atoms in each David star of DW-1 structure. Each David star is denoted by solid and dashed line (domain center and edges) and colors (domain wall) in (c).
(c) Charge characters in the energy range from 0 to 0.2 eV.
}
\end{figure}


The electronic structure of the domain wall is significantly modified from the Mott-insulator state of the C-CDW phase.
Figure 3(a) shows the total density of states (DOS) of the C-CDW phase with a Mott gap and the lower and upper Hubbard state at -0.12 eV and +0.27 eV, respectively. 
In clear contrast, the DW-1 structure exhibits nonzero DOS within the Mott gap indicating the metallic property with a shift of Hubbard states by 0.06 eV. 
As shown in more detail in Fig. 3(b), the local DOS at a domain center and edges maintain the insulating character with a consistent Mott gap as the C-CDW phase. Thus, the in-gap or metallic states are strongly confined within the domain wall, that is, on the imperfect David stars [Fig. 3(c)].


\begin{figure*}[t] 
\centering{ \includegraphics[width=16.0 cm]{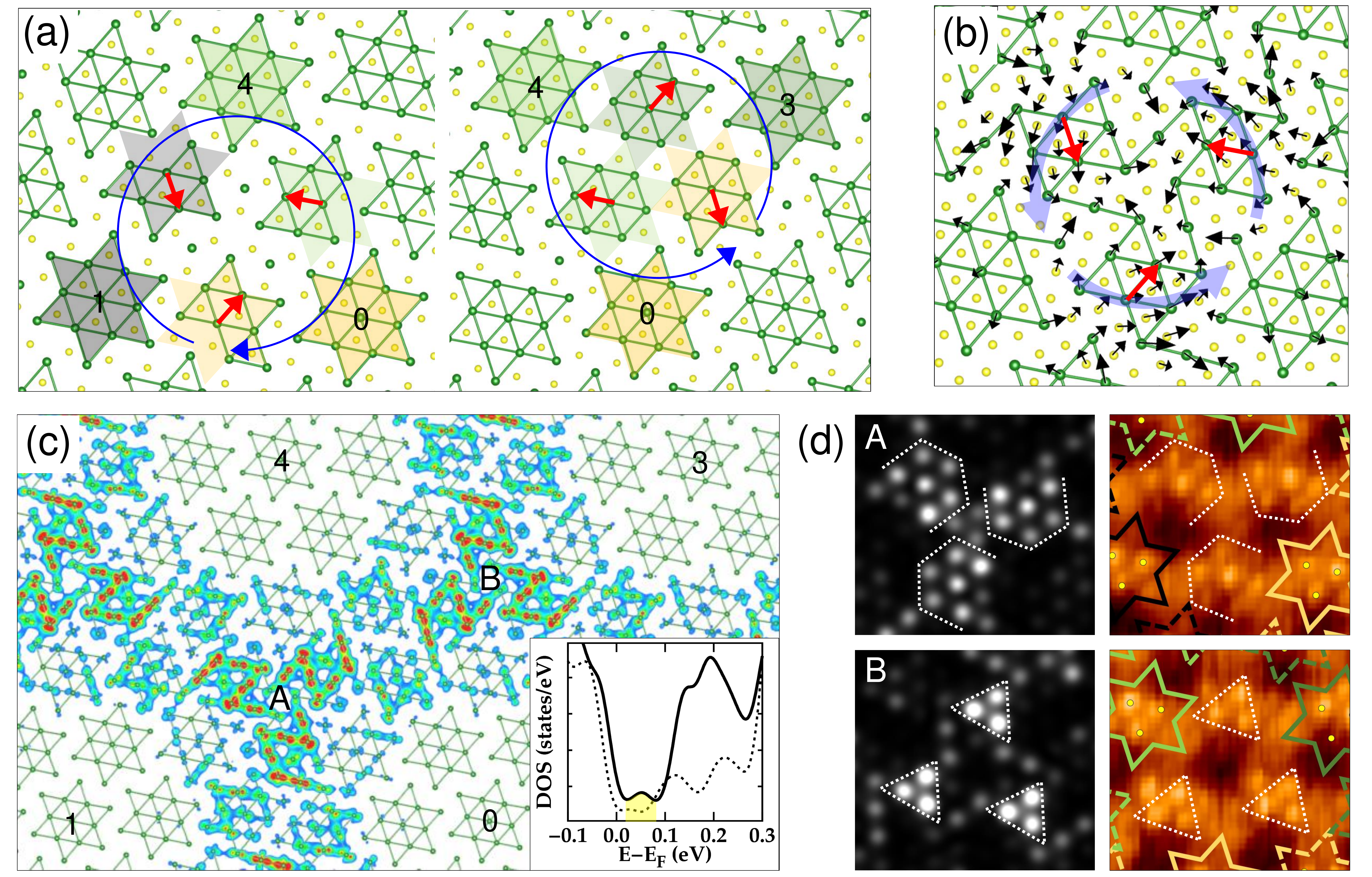} }
\caption{ \label{fig4}
(Color online)
Atomic and electronic structure of domain-wall network.
(a) Concept of the Z$_3$ vortex (clockwise, 0-1-4) and antivortex (anticlockwise, 0-3-4). Red arrows denote the phase-shift vectors between involved domains. The bottom S atoms are eliminated for better comparison of the top S atomic position at the vortex cores. 
(b) Optimized vortex core structure with atomic displacement ($\times$15). Atomic displacement is obtained from the initial state of the clockwise case in (a).  
(c) The vortex-related charge characters at Ta plane. The inset is a total DOS of the hexagonal domain model. Dashed line represents the total DOS of DW-1 for comparison. The vortex-related peak is denoted by yellow area. 
(d) Simulated and experimental STM images at vortex centers. Simulated images are obtained from the partial charge density of vortex-related state at a constant height of 2 {\AA} away from the top S atom. Experimental images in the right panel are taken from the Fig. 1(d). The dashed lines are guides for eye. The different core structures are unambiguously distinguished by the bright and dark contrast at the centers of these images.
}
\end{figure*}


We further construct the hexagonal domain lattice beyond a single DW-1 domain wall in order to describe the NC-CDW phase more closely as shown in Fig. 4.
This model has a reduced (roughly 1/3) size of the domain with about seven David-star clusters for the calculation feasibility and neighboring domains are connected by the DW-1 structure.
The total DOS of this model [inset of Fig. 4(c)] is consistent with the uniaxial DW-1 structure but much larger DOS appear in the Mott gap indicating a stronger metallic character. The extra in-gap states are due to the additional imperfect David stars in the domain-wall crossings or vortices mentioned above.
As in the case of the uniaxial domain structure, the domain regions maintain the Mott-insulating state and the metallic states are well confined to the domain walls and vortices [see Fig. 4(c) and Supplementary Materials \cite{supp17}]. 
We thus can unambiguously conclude that the NC-CDW phase is metallic and the metallic states are localized in imperfect David-star clusters of the domain-wall network.



As widely known, domain walls are topological excitations based on the degeneracy of the CDW ground state \cite{huan17}.
In the present honeycomb network, three neighboring DW-1 domain walls meet at an approximate Z$_3$ vortex or antivortex \cite{huan17}.
A vortex (antivortex) can be defined to have a clockwise (anticlockwise) rotation of phase-shift displacement vectors between involved domains [Fig. 4(a)]. 
The four domains connected by a vortex-antivortex pair should have a proper relationship of their phase shift vectors, or connected topologically, for example, when a vortex connect 0-1-4 domains, the antivortex should connect 0-3-4 domains.
After the full relaxation, a vortex and an antivortex have the same Ta atomic structure (only rotated by 180 $^{\circ}$) shown schematically in Fig. 4(b) and (c), making them electronically identical. 
Still, they have distinct structures in the S plane; one has a S atom at the center but the other does not.
The structural difference of neighboring vortex and antivortex cores in the S plane is apparent in the STM images and their simulations [Fig. 4(d)]. 
The charge analysis near the Fermi level [Fig. 4(c)] shows that the vortex (antivortex) has strong metallic electron density, which explain the extra contribution of metallic electrons in the NC-CDW phase.
The nontrivial structural property of the vortex and antivortex pairs are due basically to the large structural degrees of freedom within them with many atoms to relax.
To the best of our knowledge, this is the first case that the detailed vortex core structure of domain walls in various spin, dipole, and charge systems is disclosed.


The novel metallic domain wall network with vortex charges in the NC-CDW phase, as revealed here, can be a key to the phase transition from Mott state to superconductivity in 1T-TaS$_2$ under pressure and doping \cite{sipo08, ang12} and is consistent with the debated suggestion that the superconductivity forms within the metallic interdomain regions of the NC-CDW phase. 
We can also argue that the current picture, Mott-insulating domains with a metallic domain wall network, is valid under pressure and under the interlayer interaction. Our calculation indicates that the Mott state of C-CDW phase is robust against the in-plane (out-of-plane) lattice compression up to about 5 (6.7)$\%$.
This means that the Mott state of the domain regions of NC-CDW phase would also be robust.
On the other hand, the interlayer interaction would affect the electronic structure of the NC-CDW phase by the out-of-plane band dispersion and the change of the in-plane electronic structure by a possible stacking order \cite{rits15, rits18, guya17, lee19}.
These recent suggestions indicate metallic or small-gap insulating states, which are not consistent with the large Mott gap observed and are fragile against electron correlation [see Supplementary Materials \cite{supp17}].
The metallization scenario based on the gradual disordering of the stacking order \cite{lee19} is not compatible with the abrupt transition between NC- and C-CDW phases \cite{sipo08} and the x-ray data \cite{rits15}.
While further investigation on the interlayer interaction seems necessary, we note that there is plenty of evidence that the intralayer effect is dominant for the low temperature physics.
For example, the change of the interlayer stacking order by intrinsic domain walls in the C-CDW phase does not affect the band gap \cite{cho17}, the out-of-plane resistivity is 500 times larger than the in-plane resistivity \cite{hamb80}, and the superconducting phase is insensitive to the pressure \cite{sipo08} and the reduced dimensionality in thin flakes \cite{yu15}.


\begin{figure}[t] 
\centering{ \includegraphics[width=8.0 cm]{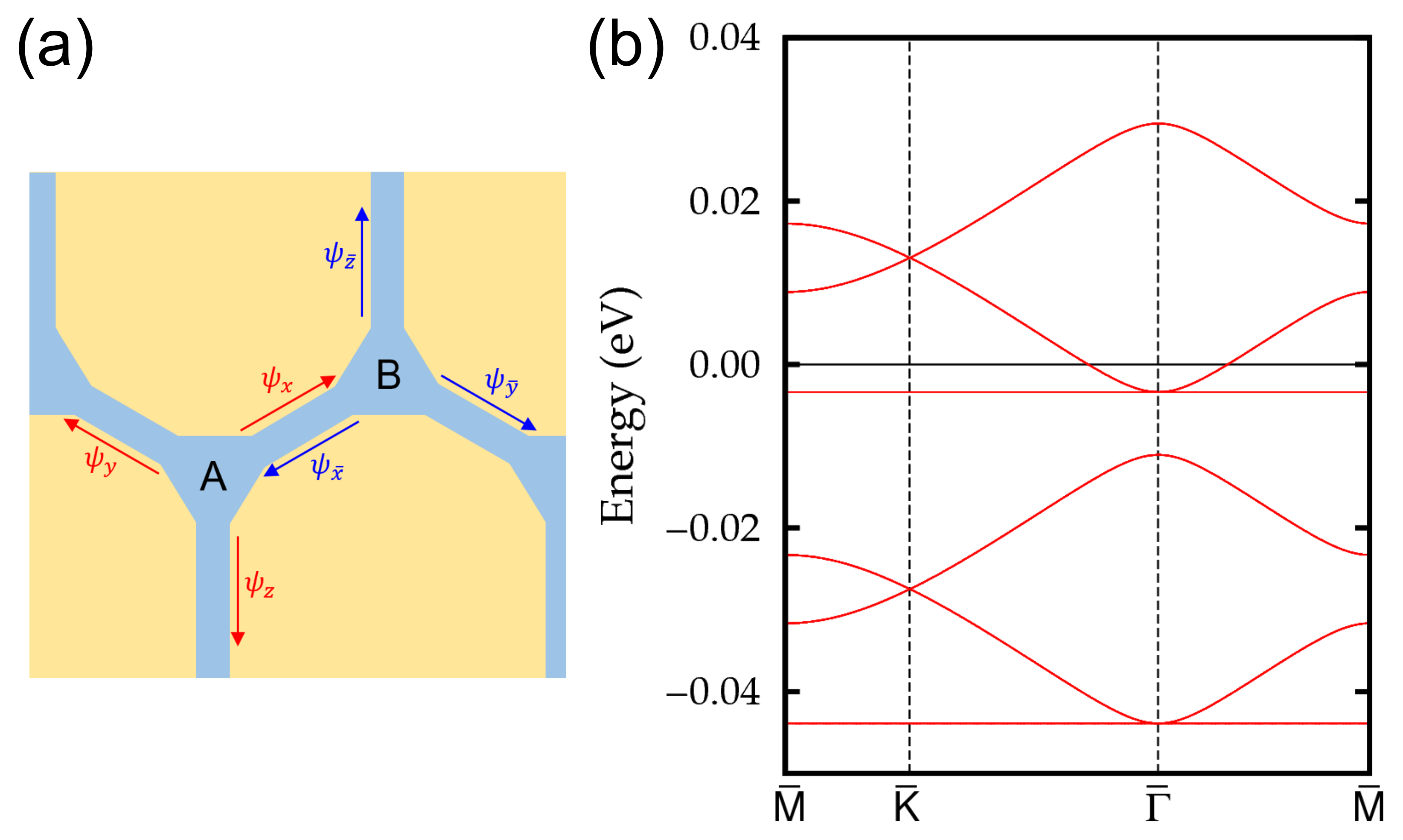} }
\caption{ \label{fig5}
(Color online)
Honeycomb lattice model of domain-wall network.
(a) Unit cell of network model. A and B denote the sublattice sites.
$\psi_a$ and $\psi_{\overline{a}}$ represent the propagating direction of the modes.
(b) Miniband structure ($v_F\hbar$ = 0.545 eV$\cdot${\AA} and $L$ = 42.24 {\AA}).
The value of $v_F\hbar$ is estimated by the band dispersion along the domain-wall direction of the DW-1 model. The same structures of minibands will repeat in energy to fill the entire Mott gap. Each one-dimensional bands inside the domain walls contribute a set of the minibands with different parameters, and hence we expect that many minibands will overlay each other in energy.
}
\end{figure}

We further emphasize that this emergent network of the domain walls and vortices has generic nature to play an essential role in many-body quantum phenomena of 1T-TaS$_2$ such as the superconducting state. 
In general, domain walls host one-dimensional metallic modes, which mix at vortices. 
Following a recent work for bilayer graphene \cite{Efimkin}, we theoretically model the system as a regular array of one-dimensional metals living on the honeycomb lattice, which is valid at the energy scale smaller than the CDW gap. 
In the ideal limit where electrons scatter only at the nodes, the system harbors a series of the nearly-flat bands with a small band width $\sim$O((2$\pi$$v$$_F$)/$L$) (here $L$ is the spatial size of the domain wall) [see Supplementary Materials \cite{supp17}]. 
The bands also feature Dirac points, flat bands, and quadratic band touchings to the flat bands as shown in Fig. 5. The former is due to the honeycomb symmetry and the latter two produce huge density of states, which would amplify any small instability of the system and enhance the tendency toward many-body correlated states including superconductivity.
Such enhancement appears  when the Fermi level is tuned near one of the cascade of the flat bands, and the tuning may be achieved by applying pressure or doping \cite{sipo08,ang12,xu10,Ang12}.
Moreover, the distinct atomic structure in the domain wall network can provides new low-energy phonon modes \cite{mcmi77}, which can favor the superconducting states as demonstrated in recent phenomenological theory \cite{chen18}. Even more interestingly, the strong correlation present in this system makes unconventional superconducting states expected theoretically \cite{chub93, nand12, lian18, po18, isob18}. 
The physics here is remarkably similar to that of twisted bilayer graphene with a small twist angle, which has a triangular network of one-dimensional metals at the domain walls between locally BA- and AB-stacked regions. Theoretically this system contains a set of nearly-flat mini-bands of huge density of states \cite{Efimkin, bist11}, similar to those of our system, being crucial for the superconductivity and Mott phases \cite{bist11, lian18, po18, isob18}. Our work extends the bilayer graphene physics to various other 2D materials but suggests beyond it with the topological degrees of freedom provided by the honeycomb nature of the network and the inherently strong electron correlation.
While the Dirac physics of the conducting honeycomb network and impact of disorders on the flat bands \cite{chal10} call for further investigation, the controllability on domain wall network suggested in recent experiments would provide exciting possibilities for engineered quantum states in this class of materials.
~\\~\\

\textbf{Methods}

The STM measurements have been carried out with a commercial microscope (Omicron) at room-temperature.
Pt-Ir wires were used as STM tips and topographic images were acquired in the constant current mode with bias voltage ($V_s$) applied to the sample. 
A single crystal of 1T-TaS$_{2}$ was prepared by iodine vapor transport method as described in Ref. \cite{cho16}.
The samples were cleaved at room temperature and quickly transferred to the precooled STM head.
DFT calculations were performed by using the Vienna $ab$ $initio$ simulation package \cite{kres96} within the Perdew-Burke-Ernzerhof generalized gradient approximation \cite{perd96} and the projector augmented wave method \cite{bloc94}. The single-layer 1T-TaS$_{2}$ was modeled with a vacuum spacing of about 13.6 {\AA}. 
We used a plane-wave basis set of 259 eV and a 6$\times$6$\times$1 $k$-point mesh for the  $\sqrt{13}\times\sqrt{13}$ unit cell and atoms were relaxed until the residual force components were within 0.05 eV/{\AA}.
To more accurately represent electronic correlations, an on-site Coulomb energy was included for Ta 5$d$ orbitals, which was estimated as 2.27 eV by the linear-response method \cite{coco05}. This value reproduces the experimental Mott gap size \cite{cho15}. To obtain the band spectrum of the honeycomb network model, we generalized the network model \cite{Efimkin} on a triangular lattice to that on the honeycomb lattice. It is equivalent to the honeycomb array of scatterers of one-dimensional metals and we solved it by using the standard Bloch wavefunction.  




This work was supported by Institute for Basic Science (Grant No. IBS-R014-D1).


\newcommand{\AP}[3]{Adv.\ Phys.\ {\bf #1}, #2 (#3)}
\newcommand{\CMS}[3]{Comput.\ Mater.\ Sci. \ {\bf #1}, #2 (#3)}
\newcommand{\PR}[3]{Phys.\ Rev.\ {\bf #1}, #2 (#3)}
\newcommand{\PRL}[3]{Phys.\ Rev.\ Lett.\ {\bf #1}, #2 (#3)}
\newcommand{\PRB}[3]{Phys.\ Rev.\ B\ {\bf #1}, #2 (#3)}
\newcommand{\NA}[3]{Nature\ {\bf #1}, #2 (#3)}
\newcommand{\NAP}[3]{Nat.\ Phys.\ {\bf #1}, #2 (#3)}
\newcommand{\NAM}[3]{Nat.\ Mater.\ {\bf #1}, #2 (#3)}
\newcommand{\NAC}[3]{Nat.\ Commun.\ {\bf #1}, #2 (#3)}
\newcommand{\NAN}[3]{Nat.\ Nanotechnol.\ {\bf #1}, #2 (#3)}
\newcommand{\NARM}[3]{Nat.\ Rev.\ Mater.\ {\bf #1}, #2 (#3)}
\newcommand{\NT}[3]{Nanotechnology {\bf #1}, #2 (#3)}
\newcommand{\JP}[3]{J.\ Phys.\ {\bf #1}, #2 (#3)}
\newcommand{\JAP}[3]{J.\ Appl.\ Phys.\ {\bf #1}, #2 (#3)}
\newcommand{\JPSJ}[3]{J.\ Phys.\ Soc.\ Jpn.\ {\bf #1}, #2 (#3)}
\newcommand{\PNAS}[3]{Proc.\ Natl.\ Acad.\ Sci. {\bf #1}, #2 (#3)}
\newcommand{\PRSL}[3]{Proc.\ R.\ Soc.\ Lond. A {\bf #1}, #2 (#3)}
\newcommand{\PBC}[3]{Physica\ B+C\ {\bf #1}, #2 (#3)}
\newcommand{\PAC}[3]{Pure Appl.\ Chem. \ {\bf #1}, #2 (#3)}
\newcommand{\SCI}[3]{Science\ {\bf #1}, #2 (#3)}
\newcommand{\SCA}[3]{Sci.\  Adv.\ {\bf #1}, #2 (#3)}
\newcommand{\RPP}[3]{Rep.\ Prog.\ Phys. \ {\bf #1}, #2 (#3)}


\begin{thebibliography}{}
\bibitem{brau04} O. M. Braun, and Y. S. Kivshar, The Frenkel-Kontorova Model : Concepts, Methods, and Applications 1st edn (Springer-Verlag, 2004).
\bibitem{fran49} F. C. Frank, and J. H. van der Merwe, One-dimensional dislocations. I. Static theory. \PRSL{A198}{205}{1949}.
\bibitem{bak82} P. Bak, Commensurate phases, incommensurate phases and the Devil{'}s staircase, \RPP{45}{587}{1982}.
\bibitem{pokr79} V. L. Pokrovsky, and A. L. Talapov, Ground-state, spectrum, and phase-diagram of 2-dimensional incommensurate crystals, \PRL{42}{65}{1979}.
\bibitem{fain80} S. C. Fain, M. D. Chinn, and R. D. Diehl, Commensurate-incommensurate transition of solid krypton monolayers on graphite, \PRB{21}{4170}{1980}.
\bibitem{rock91} A. Rockett,and C. J. Kiely, Energetics of misfit-dislocation and threading-dislocation arrays in heteroepitaxial films, \PRB{44}{1154}{1991}.
\bibitem{push02} R. Pushpa and S. Narasimhan, Stars and stripes, Nanoscale misfit dislocation patterns on surfaces, \PAC{74}{1663}{2002}.
\bibitem{chen12} S. D. Chen, Y. K. Zhou, and A. J. Soh, Molecular dynamics simulations of mechanical properties for Cu(001)/Ni(001) twist boundaries, \CMS{61}{239}{2012}.
\bibitem{wood14} C. R. Woods, L. Britnell, A. Eckmann, R. S. Ma, J. C. Lu, H. M. Guo, X. Lin, G. L. Yu, Y. Cao, R. V. Gorbachev, A. V. Kretinin, J. Park, L. A. Ponomarenko, M. I. Katsnelson, Yu. N. Gornostyrev, K. Watanabe, T. Taniguchi, C. Casiraghi, H-J. Gao, A. K. Geim and K. S. Novoselov, Commensurate-incommensurate transition in graphene on hexagonal boron nitride, \NAP{10}{451}{2014}.
\bibitem{cao18} Y. Cao, V. Fatemi, S. Fang, K. Watanabe, T. Taniguchi, E. Kaxiras, and P. Jarillo-Herrero, Unconventional superconductivity in magic-angle graphene superlattices, \NA{556}{43}{2018}.
\bibitem{wils75} J. A. Wilson, F. J. Di Salvo, and S. Mahajan, Charge density waves and superlattices in metallic layered transition-metal dichalcogenides, \AP{24}{117}{1975}.
\bibitem{mcmi76} W. L. McMillan, Theory of discommensurations and the commensurate-incommensurate charge-density-wave phase transition, \PRB{14}{1496}{1976}
\bibitem{sipo08} B. Sipos, A. F. Kusmartseva, A. Akrap, H. Berger, L. Forr{\'o}, and E. Tutis{\'s}, From Mott state to superconductivity in 1T-TaS$_2$, \NAM{7}{960}{2008}.
\bibitem{li16} L. J. Li, E. C. T. O’Farrell, K. P. Loh, G. Eda, B. {\"O}zyilmaz, and A. H. Castro Neto, Controlling many-body states by the electric-field effect in a two-dimensional material \NA{529}{185}{2016}.
\bibitem{stoj14} L. Stojchevska, I. Vaskivskyi, T. Mertelj, P. Kusar, D. Svetin, S. Brazovskii, D. Mihailovic, Ultrafast switching to a stable
hidden quantum state in an electronic crystal, \SCI{344}{177}{2014}.
\bibitem{yosh15} M. Yoshida, R. Suzuki, Y. Zhang, M. Nkano, Y. Iwasa, Memristive phase switching in two-dimensional 1T-TaS$_2$ crystals, \SCA{1}{e1500606}{2015}.
\bibitem{vask15} I. Vaskivskyi, J. Gospodaric, S. Brazovskii, D. Svetin, P. Sutar, E. Goreshnik, I. A. Mihailovic, T. Mertelj, D. Mihaiovic, Controlling the metal-to-insulator relaxation of the metastable hidden quantum state in 1T-TaS$2$, \SCA{1}{e1500168}{2015}.
\bibitem{pill00} Th. Pillo, J. Hayoz, H. Berger, R. Fasel, L. Schlapbach, and P. Aebi, Interplay between electron–electron interaction and electron–phonon coupling near the Fermi surface of 1T-TaS$_2$, \PRB{62}{4277}{2000}.
\bibitem{cler06} F. Clerc, C. Battaglia, M. Bovet, L. Despont, C. Monney, H. Cercellier, M. G. Garnier, P. Aebi, H. Berger, and L. Forr{\'o}, Lattice-distortion-enhanced electron-phonon coupling and Fermi surface nesting in 1T-TaS$_2$, \PRB{74}{155114}{2006}.
\bibitem{cho15} D. Cho, Y.-H. Cho, S.-W. Cheong, K.-S. Kim, and H. W. Yeom, Interplay of electron-electron and electron-phonon interactions in the low-temperature phase of 1T-TaS$_2$, \PRB{92}{085132}{2015}.
\bibitem{faze80} P. Fazekas and E. Tosatti, Charge carrier localization in pure and doped 1T-TaS$_2$, \PBC{99}{183}{1980}.
\bibitem{ang12} R. Ang, Y. Tanaka, E. Ieki, K. Nakayama, T. Sato, L. J. Li, W. J. Lu, Y. P. Sun, and T. Takahashi, Real-space coexistence of the melted Mott state and superconductivity in Fe-substituted 1T-TaS$_2$, \PRL{109}{176403}{2012}.
\bibitem{law17} K. T. Lawa and P. A. Lee, 1T-TaS$_2$ as a quantum spin liquid, \PNAS{114}{6996}{2017}.
\bibitem{klan17} M. Klanj{\v s}ek, A. Zorko, R. {\v Z}itko, J. Mravlje, Z. Jaglicic, P. K. Biswas, P. Prelov{\v s}ek1,5, D. Mihailovic and D. Arcon, A high-temperature quantum spin liquid with polaron spins, \NAP{13}{1130}{2017}.
\bibitem{brou80} R. Brouwer and F. Jellinek, The low-temperature superstructures of 1T-TaSe$_2$ and 2H-TaSe$_2$, \PBC{99}{51}{1980}.
\bibitem{wu89} X. L. Wu and C. M. Lieber, Hexagonal domain-like charge density wave phase of TaS$_2$ determined by scanning tunneling microscopy, \SCI{243}{1704}{1989}.
\bibitem{naka77} K.Nakanishi, H. Takatera, Y. Yamada, and H. Shiba, The nearly commensurate phase and effect of harmonics on the successive phase transition in 1T-TaS$_2$, \JPSJ{43}{1509}{1977}.
\bibitem{yama83} A. Yamamoto, Hexagonal domainlike structure in 1T-TaS$_2$, \PRB{27}{7823}{1983}.
\bibitem{wu90} X. L. Wu and C. M. Lieber, Direct observation of growth and melting of the hexagonal-domain charge-density-wave phase in 1 T-TaS$_2$ by scanning tunneling microscopy, \PRL{64}{1150}{1990}.
\bibitem{burk91} B. Burk, R. E. Thomson, A. Zettl, and J. Clarke, Charge-density-wave domains in 1T-TaS$_2$ observed by satellite structure in scanning- tunneling-microscopy images, \PRL{66}{3040}{1991}.
\bibitem{spij97} A. Spijkerman, J. L. de Boer, A. Meetsma, and G. A. Wiegers, X-ray crystal-structure refinement of the nearly commensurate phase of 1T-TaS$_2$ in (3+2)-dimensional superspace, \PRB{56}{1357}{1997}.
\bibitem{tsen15} A. W. Tsen, et al. Structure and control of charge density waves in two-dimensional 1T-TaS$_2$, \PNAS{112}{15054}{2015}.
\bibitem{cho16} D. Cho,  S. Cheon, K.-S. Kim, S.-H. Lee, Y.-H. Cho, S.-W. Cheong, and H. W. Yeom, Nanoscale manipulation of the Mott insulating state coupled to charge order in 1T-TaS$_2$, \NAC{7}{10453}{2016}.
\bibitem{ma16} L. Ma, C. Ye, Y. Yu, X. F. Lu, X. Niu, S. Kim, D. Feng, D Tom{\'a}nek, Y.-W. Son, X, H, Chen, and Y. Zhang, A metallic mosaic phase and the origin of Mott insulating state in 1T-TaS$_2$. \NAC{7}{10956}{2016}.
\bibitem{vask16} I. Vaskivskyi,  I. A. Mihailovic, S. Brazovskii, J. Gospodaric, T. Mertelj, D. Svetin, P. Sutar, and D. Mihailovic, Fast electronic resistance switching involving hidden charge density wave states, \NAC{7}{11442}{2016}.
\bibitem{xu10} P. Xu, J. O. Piatek, P.-H. Lin, B. Sipos, H. Berger, L. Forr{\'o}, H. M. R$\phi$nnow, and M. Grioni, Superconducting phase in the layered dichalcogenide 1T-TaS$_2$ upon inhibition of the metal-insulator transition, \PRB{81}{172503}{2010}.
\bibitem{Ang12} R. Ang, R. Ang, Y. Miyata, E. Ieki, K. Nakayama, T. Sato, Y. Liu, W. J. Lu, Y. P. Sun, and T. Takahashi Superconductivity and bandwidth-controlled Mott metal–insulator transition in 1T-TaS$_{2-x}$Se$_x$, \PRB{88}{115145}{2013}.
\bibitem{yu15} Y. Yu, F. Yang, X. F. Lu, Y. J. Yan, Y.-H Cho, L. Ma, X. Niu, S. Kim, Y.-W. Son, D. Feng, S. Li, S.-W. Cheong, X. H. Chen, and Y. Zhang, Gate-tunable phase transitions in thin flakes of 1T-TaS$2$, \NAN{10}{270}{2015}.
\bibitem{liu16} Y. Liu, D. F. Shao, L. J. Li, W. J. Lu, X. D. Zhu, P. Tong, R. C. Xiao, L. S. Ling, C. Y. Xi, L. Pi, H. F. Tian, H. X. Yang, J. Q. Li, W. H. Song, X. B. Zhu, and Y. P. Sun, Nature of charge density waves and superconductivity in 1T-TaSe$_{2-x}$Te$_x$, \PRB{94}{045131}{2016}.
\bibitem{cho17} D. Cho, G. Gye, J. Lee, S.-H. Lee, L. Wang, S.-W. Cheong, and H. W. Yeom, Correlated electronic states at domain walls of a Mott-charge-density-wave insulator 1T-TaS$_2$, \NAC{8}{392}{2017}.
\bibitem{rits13} T. Ritschel, J. Trinckauf, G. Garbarino, M. Hanﬂand, M. v. Zimmermann, H. Berger, B. B{\"u}chner, and J. Geck, Pressure dependence of the charge density wave in 1T-TaS$_2$ and its relation to superconductivity, \PRB{87}{125135}{2013}.
\bibitem{hovd16} R. Hovdena, A. W. Tsen, P. Liua, B. H. Savitzkye, I. El Baggarie, Y. Liuf, W. Luf, Y. Sun, P. Kimi, A. N. Pasupathy, and Lena F. Kourkoutis, Atomic lattice disorder in charge-density-wave phases of exfoliated dichalcogenides (1T-TaS$_2$) \PNAS{113}{11420}{2016}.
\bibitem{ishi91} T. Ishiguro and H. Sato, Electron microscopy of phase transformations in 1T-TaS$_2$, \PRB{44}{2046}{1991}.
\bibitem{han15} T.-R. Han, F. Zhou, C. D. Malliakas, P. M. Duxbury, S. D. Mahaniti, M. G. Kanatzidis, C.-Y. Ruan, Exploration of metastability and hidden phase in correlated electron crystals visualized by femtosecond optical doping and electron crystallography, \SCA{1}{e1400173}{2015}.
\bibitem{supp17} Supplementary Materials
\bibitem{rits15} T. Ritschel, J. Trinckauf, K. Koepernik, B. B{\"u}chner, M. v. Zimmermann, H. Berger, Y. I. Joe, P. Abbamonte, and J. Geck, Orbital textures and charge density waves in transition metal dichalcogenides, \NAP{11}{328}{2015}.
\bibitem{rits18} T. Ritschel, H. Berger, and J. Geck, Stacking-driven gap formation in layered 1T-TaS$_2$, \PRB{98}{195134}{2018}.
\bibitem{guya17} L. Le Guyader, T. Chase, A. H. Reid, R. K. Li, D. Svetin, X. Shen, T. Vecchione, X. J. Wang, D. Mihailovic, and H. A. D{\"u}rr, Stacking order dynamics in the quasi-two-dimensional dichalcogenide 1T-TaS$_2$ probed with MeV ultrafast electron diffraction, Struct. Dyn. {\bf 4}, 044020 (2017).
\bibitem{lee19} S.-H. Lee, J. S. Goh, and D. Cho, Origin of the insulating phase and first-order metal-insulator transition in 1T-TaS$_2$, \PRL{122}{106404}{2019}.
\bibitem{hamb80} P.D. Hambourger and F.J. Di Salvo, Electronic conduction process in 1T-TaS$_2$, \PRB{99B}{173}{1980}.
\bibitem{huan17} F.-T. Huang and S.-W. Cheng, Aperiodic topolgical order in the domain configurations of functional material, \NARM{2}{17004}{2017}.
\bibitem{Efimkin} D. K. Efimkin and A. H. MacDonald, Helical network model for twisted bilayer graphene, \PRB{98}{035404}{2018}.
\bibitem{mcmi77} W. L. McMillan, Collective modes of a charge-density wave near the lock-in transition, \PRB{16}{4655}{1977}.
\bibitem{chen18} C. Chen, L. Su, A. H. Castro Neto, Vitor M. Pereira, Discommensuration-enhanced superconductivity in the charge density wave phases of transition-metal dichalcogenides, ArXiv e-print (2018), arXiv:1806.08064 [cond-mat.mes-hall].
\bibitem{chub93} Andrey V. Chubukov, Kohn-Luttinger effect and the instability of a two-dimensional repulsive Fermi liquid at T=0, \PRB{48}{1097}{1993}.
\bibitem{nand12} R. Nandkishore, L. S. Levitov, and A. V. Chubukov, Chiral superconductivity from repulsive interactions in doped graphene, \NAP{8}{158}{2012}.
\bibitem{lian18} B. Lian, Z. Wang, B. A. Berneving, Twisted Bilayer Graphene: A Phonon Driven Superconductor, ArXiv e-print (2018), arXiv:1807.04382 [cond-mat.mes-hall].
\bibitem{po18} H. C. Po, L. Zou, A. Vishwanath, and T. Senthil, Origin of Mott insulating behavior and superconductivity in twisted bilayer graphene, ArXiv e-print (2018), arXiv:1803.09742 [cond-mat.str-el].
\bibitem{isob18} H. Isobe, N. F. Q. Yuan, and L. Fu, Unconventional superconductivity and density waves in twisted bilayer graphene, ArXiv e-print (2018), arXiv:1805.06449 [cond-mat.str-el].
\bibitem{bist11} R. Bistritzer and A. H. MacDonald, Moir{\'e} bands in twisted double-layer graphene, \PNAS{108}{12233}{2011}.
\bibitem{chal10} J. T. Chalker, T. S. Pickles, and Pragya Shukla, Anderson localization in tight-binding models with flat bands, \PRB{82}{104209}{2010}. 
\bibitem{kres96} G. Kresse and J. Furthm{\"u}ller, Efficient iterative schemes for $ab$ $initio$ total-energy calculations using a plane-wave basis set, \PRB{54}{11169}{1996}.
\bibitem{perd96} J. P. Perdew, K. Burke, and M. Ernzerhof, Gerneralized gradient approximation made simple, \PRL{77}{3865}{1996}.
\bibitem{bloc94} P. E. Blochl, Projector augmented-wave method, \PRB{50}{17953}{1994}.
\bibitem{coco05} M. Cococcioni and S. de Gironcoli, Linear response approach to the calculation of the effective interaction parameters in the LDA+U method, \PRB{71}{0351058}{2005}.



\end{thebibliography}
\end{document}


\title{Supplementary Materials for ``Emergent Network of Topological Excitations in Correlated Charge Density Wave"}
\author{Jae Whan Park$^1$}
\author{Gil Young Cho$^2$}
\author{Jinwon Lee$^{1,2}$}
\author{Han Woong Yeom$^{1,2}$}
\email{yeom@postech.ac.kr}
\affiliation{$^1$Center for Artificial Low Dimensional Electronic Systems,  Institute for Basic Science (IBS),  77 Cheongam-Ro, Pohang 790-7884, Korea.} 
\affiliation{$^2$Department of Physics, Pohang University of Science and Technology, Pohang 790-784, Korea}

\date{\today}
\maketitle
\appendix

\section{\fontsize{10}{15}\selectfont A. Details of STS spectra and DFT calculations}

\textbf{We provide additional information about the STS spectra, Mott states of the theoretical models and interlayer coupling effect.}

\renewcommand{\figurename}{SFIG.}

\begin{figure} [h!]
\centering{ \includegraphics[width=14.5cm]{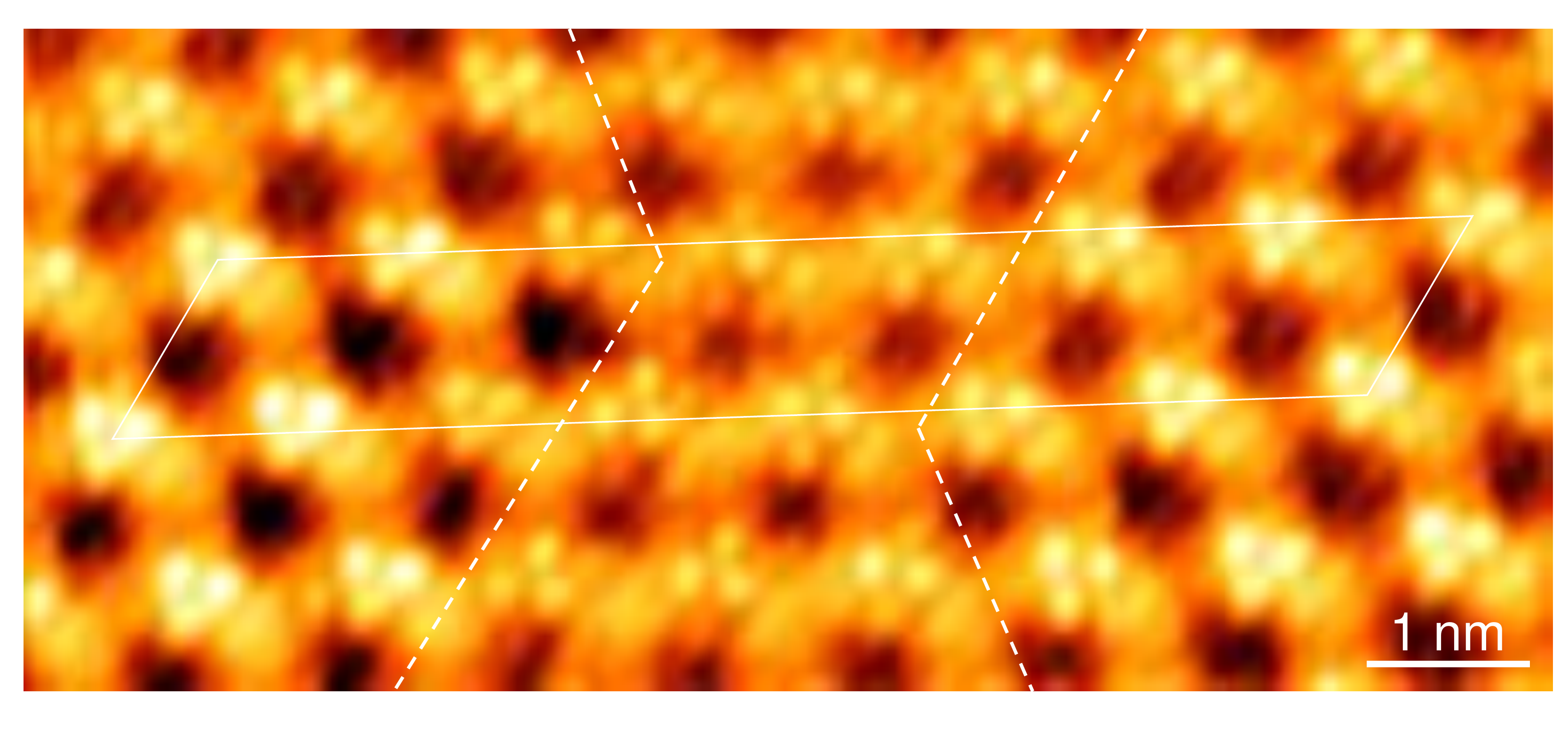} }
\caption{ \label{fig1}
Well filtered STM image of the NC phase.
Dashed lines separate two different domains and domain wall.
The solid lines represent the same area as Fig. 2(c).
}
\end{figure}

\begin{figure} [h!]
\centering{ \includegraphics[width=16cm]{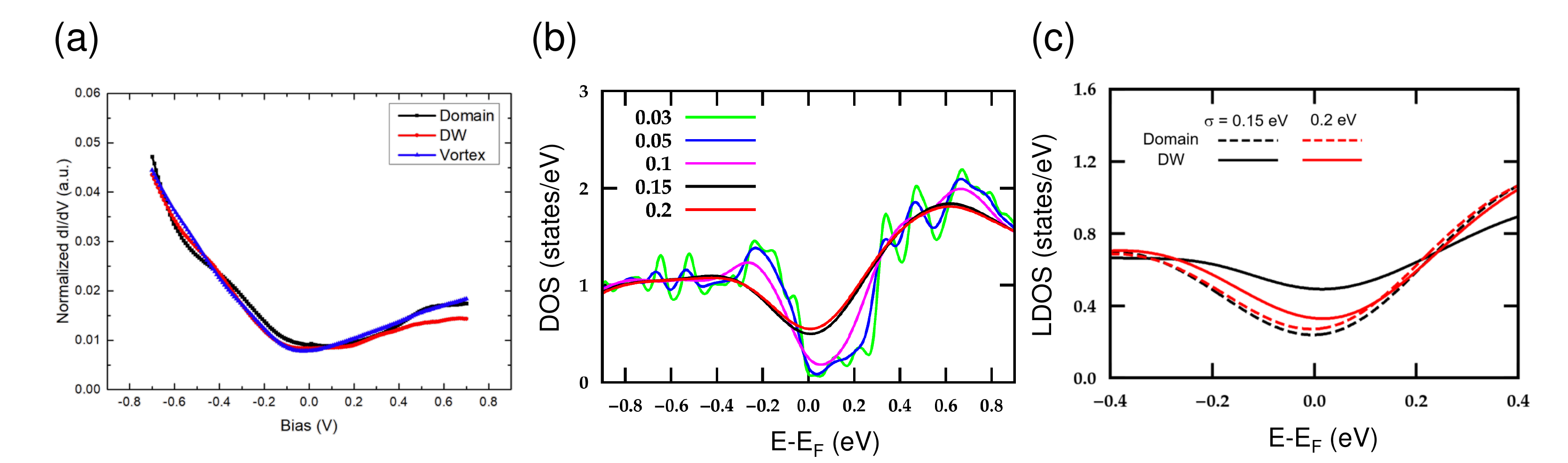} }
\caption{ \label{fig1}
STS spectra of the NC phase and theoretical DOS of DW-1 structure. (a) STS spectra, (b) Total DOS as a function of the electronic temperature parameter $\sigma$ (eV). (c) Local DOS of domain and domain wall regions ($\sigma$ = 0.15 and 0.2 eV). 
}
\end{figure}

\begin{figure} [ht]
\centering{ \includegraphics[width=16cm]{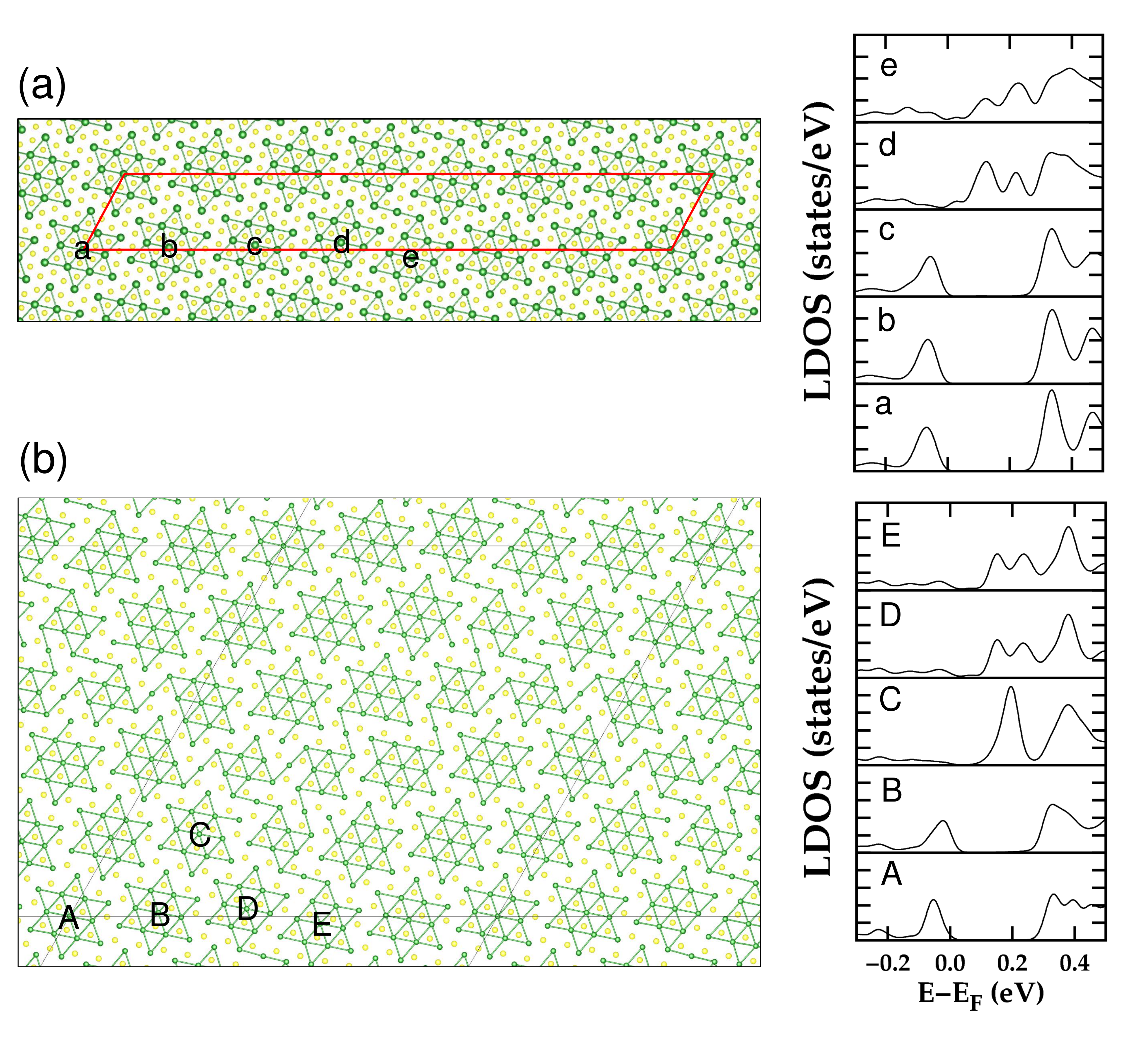} }
\caption{ \label{fig2}
Local density of states at center Ta atom of David stars (a) DW-1 and (b) Hexagonal domain model.
}
\end{figure}

\clearpage

\begin{figure} [h!]
\centering{ \includegraphics[width=16cm]{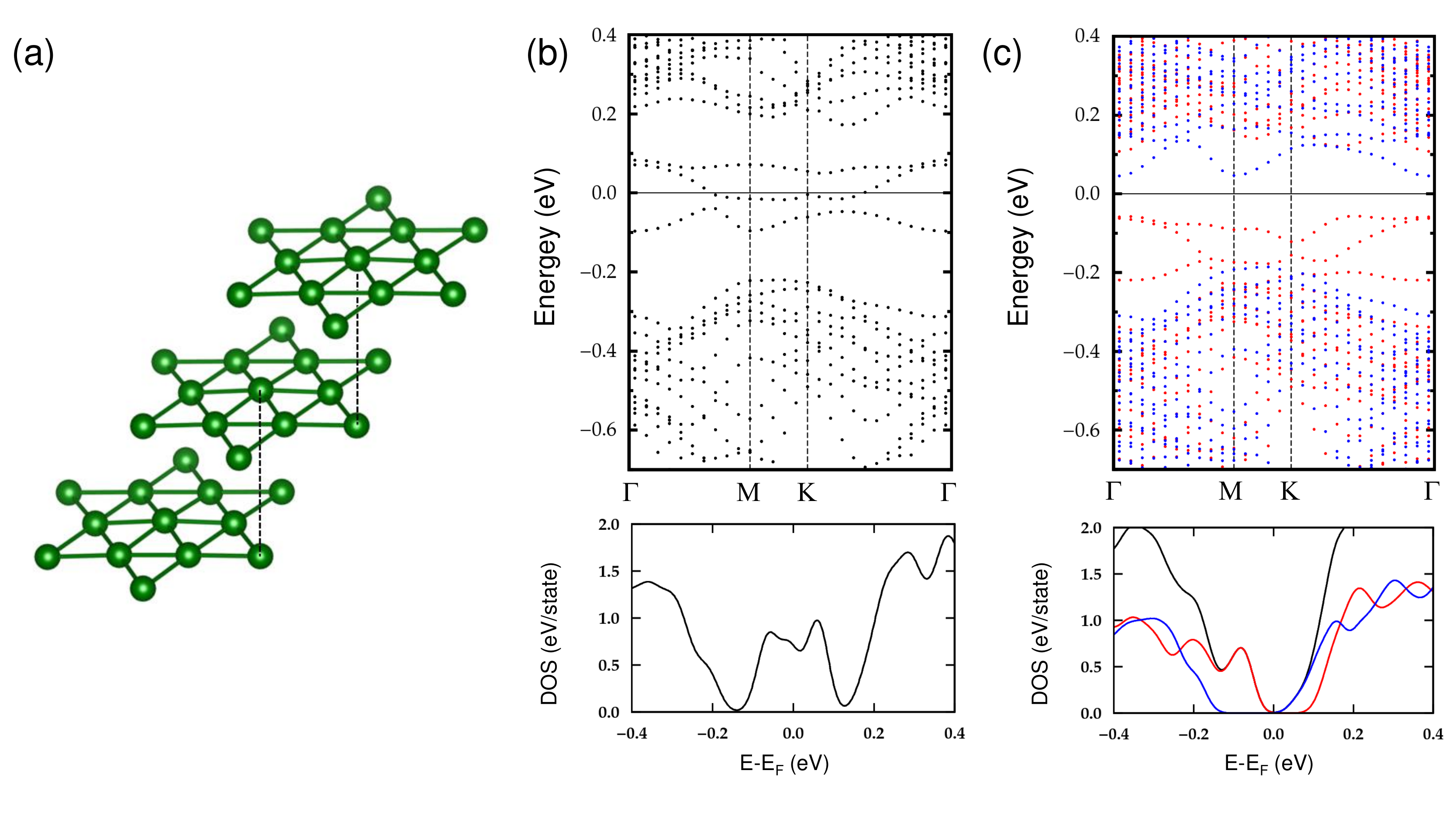} }
\caption{ \label{fig3}
One possible tri-layer stacking order in the C phase. (a) atomic structure, (b) and (c) are the band structure and total DOS without and with electron-electron correlations, respectively. The interlayer distance of 5.9 {\AA} is fixed.
}
\end{figure}

\begin{figure} [h!]
\centering{ \includegraphics[width=16cm]{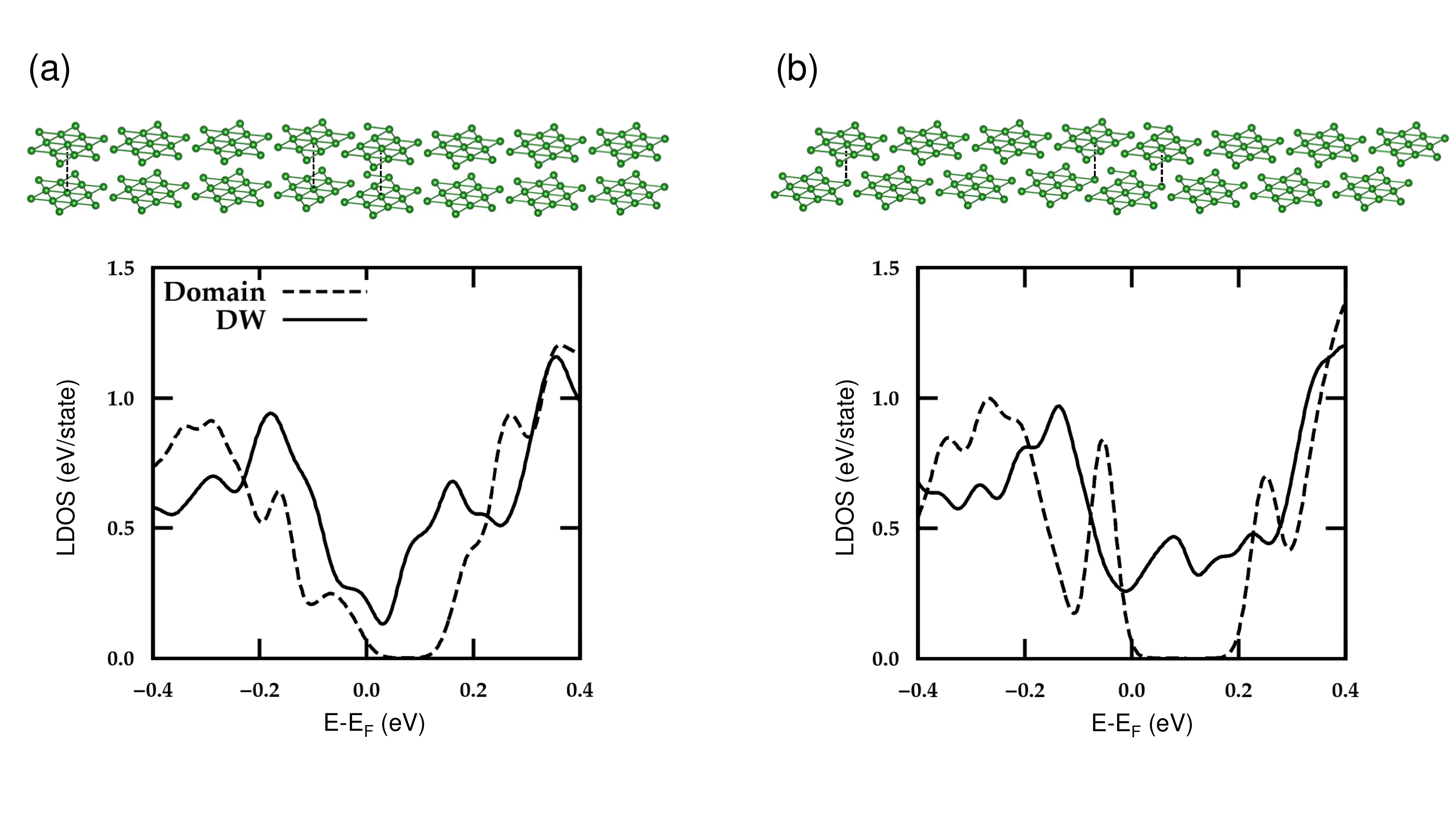} }
\caption{ \label{fig4}
Two different bilayer stacking order in the DW-1 structure. (a) on-top stacking and (b) shifted stacking. Top and bottom panels are atomic structure and LDOS, respectively. All atoms are fixed at their atomic position in the single-layer DW-1 structure. The interlayer distance of 5.9 {\AA} is fixed.
}
\end{figure}

\clearpage

\section{\fontsize{10}{15}\selectfont B. Details of Honeycomb Network Model}
In this supplementary material, we present the details of the honeycomb network model. Here we apply the strategy of Ref. \cite{Efimkin} to the honeycomb lattice.

\subsection{Construction of Network Model}

We theoretically model the system by a regular array of one-dimensional metals living on the links of a honeycomb lattice. The construction of this model is motivated from the following experimental observations present in the main text. 
\begin{enumerate}
\item \textbf{Emergent honeycomb lattice}: The domains of the NC-CDW state form a regular honeycomb lattice, and the domain walls are the links of this honeycomb lattice.
\item \textbf{Metallic domain walls}: The domain walls trap finite local density of states near the Fermi level (Note that this is generically expected for any domain wall of a charge-density wave since the domain wall carries in-gap states whose origin are topological \cite{su79}).   
\end{enumerate} 

Motivated from these, we consider a regular array of one-dimensional metals living on the links of a honeycomb lattice. Similar network models of one-dimensional metals have been studied to some degree in the context of quantum Hall plateau transitions, known as ``Chalker-Coddington model" \cite{chal88}, and recently have been revived to explain the physics of the twisted bilayer graphene at a small twisting angle \cite{Efimkin}. We apply this latest theoretical progress to model the network on the honeycomb lattice. 

Some details of the model are following:  
\begin{enumerate}
\item \textbf{Degrees of freedom}: On each link $a=x,y,z$ of the honeycomb lattice, we assign the two wavefunctions $\psi_a$ and $\psi_{\bar{a}}$. Here $\psi_a$ represent the chiral mode propagating from an A-sublattice to its neighboring B-sublattice and $\psi_{\bar{a}}$ for the mode propagating from a B-sublattice to its neighboring A-sublattice (See Fig. \ref{fig1}). 
Microscopically they correspond to the low-energy modes near the Fermi momentum of one-dimensional metals propagating along the links. (Here we suppress the spin index for the modes because we are mainly interested in the spectral properties of the network model.)
\item \textbf{Scattering between wavefunctions}: We assume that the modes propagate coherently within each link and scatter only at the nodes of the honeycomb lattice. We further assume that there are three-fold rotation and two-fold mirror symmetries at each nodes, and the scattering between the modes respects the crystal symmetries. 
\end{enumerate}

\begin{figure}[htbp]
\centering\includegraphics[width=0.5\textwidth]{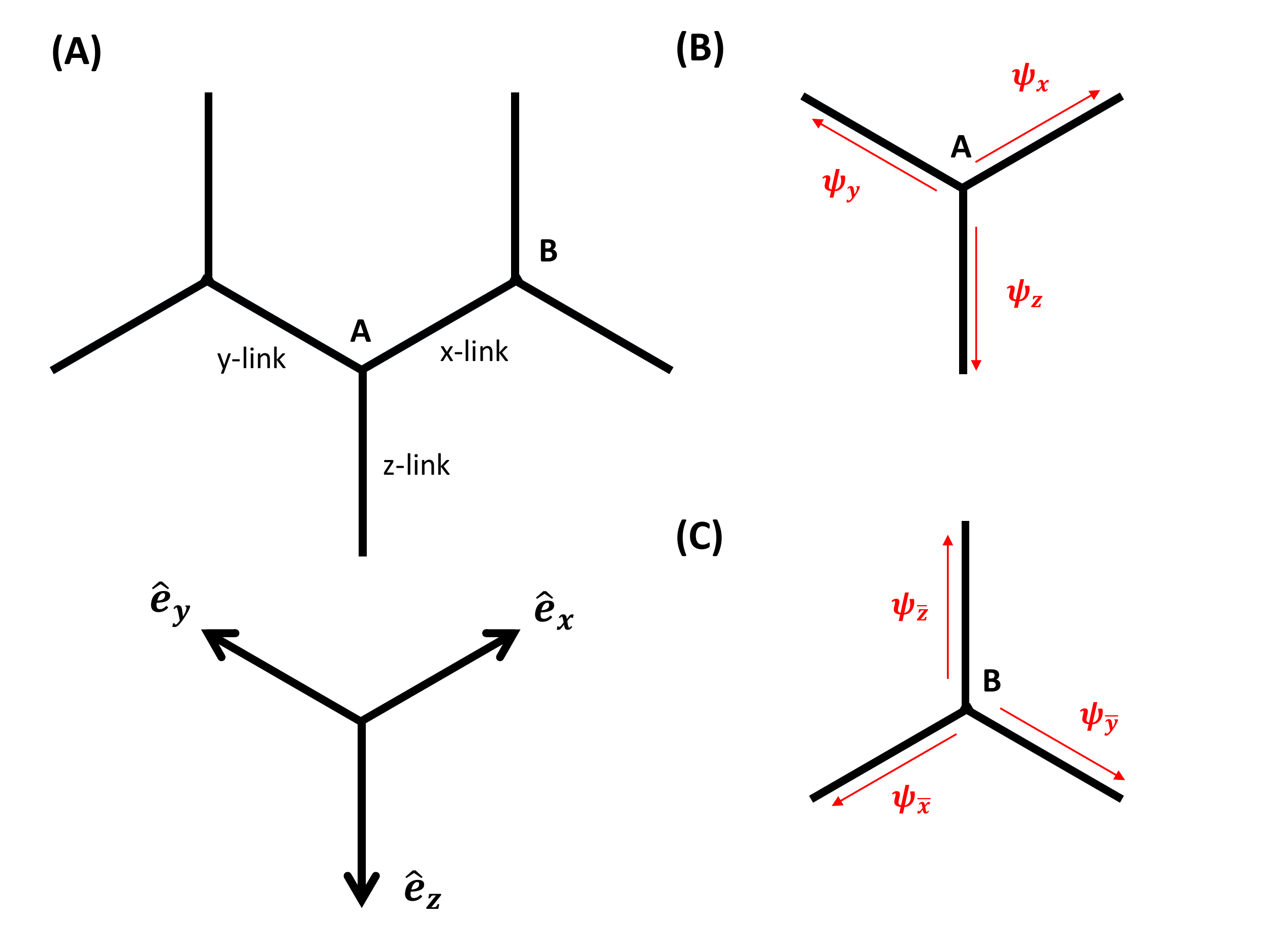}
\caption{Pictorial Representation of Network Model: (A) labeling of the links by $x,y,z$ on real space. Each link has the two degrees of freedom. One is from the neighboring A-sublattice [See (B)] and another is from the neighboring B-sublattice [See (C)]. They are written as $\psi_a$ and $\psi_{\bar{a}}$, and the arrows represent the propagating directions of the modes. Here $\hat{e}_a, a= x,y,z$ is the vector connecting the A-sublattice to the B-sublattice.}\label{fig1}
\end{figure}

Having these in mind, we have the following scattering problem at the A-sublattice.   
\begin{align}
\left[
\begin{array}{c}
\psi_x (\bm{R}) \\
\psi_y (\bm{R}) \\ 
\psi_z (\bm{R}) 
\end{array} \right] = e^{-i\frac{E}{v_F \hbar} L} \cdot \hat{T}_A \cdot \left[
\begin{array}{c}
\psi_{\bar{x}} (\bm{R} +\hat{e}_x) \\
\psi_{\bar{y}}  (\bm{R} + \hat{e}_y) \\ 
\psi_{\bar{z}}  (\bm{R} +\hat{e}_z) 
\end{array} \right]
\end{align}
Here, the left-hand side $\psi_a (\bm{R}), a= x,y,z$ represents the out-going modes from the A-sublattice, which is related by a scattering matrix $\hat{T}_A$ to the in-coming modes $\psi_{\bar{a}} (\bm{R}), a = x,y,z$ appearing on the right-hand side (See Fig. \ref{fig1}). The additional phase factor $\sim \exp (-i\frac{E}{v_F \hbar} L)$ is the phase accumulated by the incoming modes while it propagates coherently from the neighboring B-sublattices to the A-sublattice at $\bm{R}$. Here $v_F$ is the Fermi velocity within the one-dimensional metal, which is expected to be similar to that of the bulk electron, and $L$ is the length of the link. The scattering matrix $\hat{T}_A$ can be fixed by the three-fold rotation as well as the two-fold mirrors at the node. With the unitarity of the scattering matrix, we find 
\begin{align}
\hat{T}_A = e^{i\chi_A} \left[
\begin{array}{ccc}
T_A & t_A & t_A \\
t_A & T_A & t_A \\ 
t_A & t_A & T_A  
\end{array} \right], ~~ |T_A| \in \Big[\frac{1}{3}, ~1\Big], ~ t_A   = e^{i\phi_A} \sqrt{\frac{1-|T_A|^2}{2}},  
\end{align}
with $\phi_A = \cos^{-1}(\frac{|t_A|}{2|T_A|})$. Similarly we have the following scattering problem at the B-sublattice.
\begin{align}
\left[
\begin{array}{c}
\psi_{\bar{x}} (\bm{R}) \\
\psi_{\bar{y}} (\bm{R}) \\ 
\psi_{\bar{z}} (\bm{R}) 
\end{array} \right] = e^{-i\frac{E}{v_F \hbar} L} \cdot \hat{T}_B \cdot \left[
\begin{array}{c}
\psi_{x} (\bm{R} -\hat{e}_x) \\
\psi_{y}  (\bm{R} - \hat{e}_y) \\ 
\psi_{z}  (\bm{R} -\hat{e}_z) 
\end{array} \right],
\end{align}
where $\hat{T}_B$ has the same structure as the $\hat{T}_A$. 

Now we can perform the Fourier transformation on $\bm{R}$ and solve these scattering problems. 
\begin{align}
\bm{\Psi}_{\bm{q}} = e^{-i\frac{E_{\bm{q}}}{v_F \hbar} L} \hat{T}_{\bm{q}} \cdot \bm{\Psi}_{\bm{q}}, ~~ \bm{\Psi}_{\bm{q}} = \left[
\begin{array}{c}
\psi_{x} (\bm{q}) \\
\psi_{y} (\bm{q}) \\ 
\psi_{z} (\bm{q}) \\
\psi_{\bar{x}} (\bm{q}) \\
\psi_{\bar{y}} (\bm{q}) \\ 
\psi_{\bar{z}} (\bm{q})  
\end{array} \right], ~~ 
\hat{T}_{\bm{q}} = \left[
\begin{array}{cc}
0 & \hat{T}_A \cdot \hat{V}_{\bm{q}} \\ 
\hat{T}_B \cdot \hat{V}_{\bm{q}}^* & 0  
\end{array} \right],
\label{Energy}
\end{align}
where $\hat{V}_{\bm{q}} = $ diag $[\exp (i\bm{q}\cdot \hat{e}_x), ~\exp (i\bm{q}\cdot \hat{e}_y), ~\exp (i\bm{q}\cdot \hat{e}_z)]$. Hence, the energy spectrum can be obtained by diagonalizing $\hat{T}_{\bm{q}}$, which is again an unitary matrix. In terms of the eigenvalues $e^{i \epsilon_{j} (\bm{q})}, j= 1, 2, \cdots 6$ of $\hat{T}_{\bm{q}}$, we have the energy spectrum: 
\begin{align}
E_{j, \bm{q}}^n = 2\pi \frac{v_F \hbar}{L} n + \frac{v_F \hbar}{L} \epsilon_{j} (\bm{q}), ~~j = 1,2, \cdots 6.
\end{align}
Here $n \in \mathbb{Z}$ and thus the minibands are repeating in the energy in period of $2\pi \frac{v_F \hbar}{L}$. Mathematically this repetition in $n$ originates from the ambiguity of $\epsilon_j (\bm{q})$ by $2\pi$ appearing in the eigenvalues $e^{i \epsilon_{j} (\bm{q})}, j= 1, 2, \cdots 6$. Physically this repetition can be traced back to the excitation energy of the microscopic one-dimensional modes with the same momentum $\bm{q}$, i.e., for a given $\bm{q}$, there are multiple different one-dimensional modes with energy $2\pi \frac{v_F \hbar}{L} n, ~ n\in \mathbb{Z}$. Thus we expect that the energy spectrum given by $\frac{v_F \hbar}{L}\epsilon_j (\bm{q})$ will repeat in energy with a period $2\pi \frac{v_F \hbar}{L}$ and entirely fills up the bulk CDW gap. Below we will analyze only one period of the band spectrum. 

\subsection{Band Spectrum}
We first consider the case where we have a full symmetry of the honeycomb lattice, i.e., $\hat{T}_A = \hat{T}_B$. As apparent from the Fig \ref{fig2}, the spectrum features (i) Dirac cones at the $K$ and $K'$ points, (ii) flat bands, and (iii) quadratic band touchings at the $\Gamma$ point. Now we discuss the stabilities of these features. 
\begin{enumerate}
\item \textbf{Dirac Cones at the $K$ and $K'$ points}: The Dirac cones are protected by the sublattice symmetry as in the graphene. It is easily removed by breaking the symmetry, i.e., $T_A \neq T_B$. See the spectrum in Fig \ref{fig2}.
\item \textbf{Quadratic Band Touching at the $\Gamma$ point}: The quadratic band touchings can be protected by the six-fold rotation symmetry \cite{sun09}. However, even when the symmetry is broken (while keeping the three-fold rotation and mirror symmetries are kept), the band touchings are robust within our network model. See the Fig \ref{fig2}.  
\item \textbf{``Flat-ness" of Flat bands}: The flat-ness of the bands cannot be protected. However, within our network model (with the three-fold rotation $C_3$ and mirror symmetries), we find that it is robust. See the Fig \ref{fig2}. 
\end{enumerate}

\begin{figure}[htbp]
\centering\includegraphics[width=0.5\textwidth]{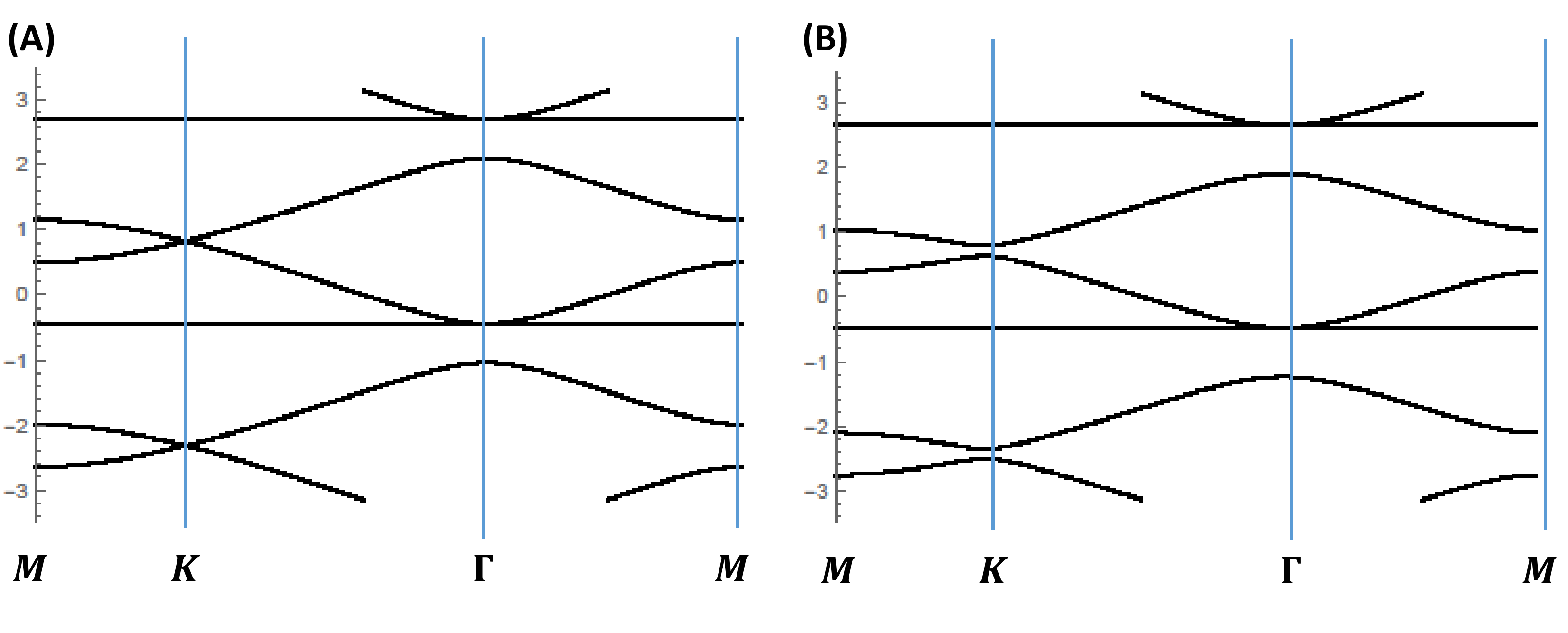}
\caption{Energy Spectrum of Honeycomb Network Model. Here we plot the spectrum of $\epsilon_j (\bm{q})$ of the scattering matrix $\hat{T}_{\bm{q}}$ along $M \to K \to \Gamma \to M$. (A) $C_6$-symmetric spectrum $\hat{T}_A = \hat{T}_B$. It is straightforward to note the Dirac fermion at the $K$ point, quadratic band touching at $\Gamma$ point, and also the flat bands. (B) $C_3$-symmetric spectrum $\hat{T}_A \neq \hat{T}_B$. Here the Dirac band touching is removed.}\label{fig2}
\end{figure}

\subsection{Symmetry Analysis of Quadratic Band Touching}  
Here we perform the symmetry analysis of the quadratic band touchings at the $\Gamma$ point. By explicitly diagonalizing Eq.\eqref{Energy} at the $\Gamma$ point (with $\hat{T}_A = \hat{T}_B$), we obtain the wavefunctions of the degenerate states. They are labeled as $|\Psi_a \rangle, a =1,2$ (with $\langle \Psi_a | \Psi_b \rangle = \delta_{ab}$), from which we reconstruct the representations of the three-fold rotation $C_3$, the x-mirror $R_x: x \to -x$, and the six-fold rotation $C_6$, i.e., $\{[C_3]_p, [R_x]_p, [C_6]_p\}$ within these bands. For example, we obtain the three-fold rotation $[C_3]_p$ within the two states by computing
\begin{align}\label{Ex}
[C_3]_p = \left[
\begin{array}{cc}
\langle \Psi_1 | \hat{C}_3 |\Psi_1 \rangle & \langle \Psi_2 | \hat{C}_3 |\Psi_1 \rangle \\ 
\langle \Psi_1 | \hat{C}_3 |\Psi_2 \rangle & \langle \Psi_2 | \hat{C}_3 |\Psi_2 \rangle
\end{array} \right],
\end{align} 
in which $\hat{C}_3$ is the representation of the three-fold rotation in the six-component $\Psi_{\bm{q}}$ in Eq.\eqref{Energy}. 




By computing these explicitly, we find
\begin{align}
[C_3]_p = e^{\frac{2\pi i}{3} \sigma^z}, ~ [R_x]_p = e^{-\frac{\pi i}{3} \sigma^z} \sigma^x, ~[C_6]_p = e^{-\frac{2\pi i}{3} \sigma^z}, 
\end{align}
where $\sigma^a, a= x,y,z$ is the Pauli matrix acting on the space spanned by $\{|\Psi_1 \rangle, |\Psi_2 \rangle\}$.

With these, we can write down symmetry-allowed Hamiltonian near the $\Gamma$ point. First, by imposing $[C_3]_p$ and $[R_x]_p$, we find that no perturbation is allowed to split $|\Psi_1 \rangle$ and $|\Psi_2 \rangle$ at the $\Gamma$ point.  
\begin{align}
[C_3]_p^{\dagger} H_0 [C_3]_p = H_0, ~~[R_x]_p^{\dagger} H_0 [R_x]_p = H_0, ~ \to H_0 \propto \mu \sigma^0
\end{align}
Hence the degeneracy cannot be removed when $[C_3]_p$ and $[R_x]_p$ are imposed. On the other hand, near the $\Gamma$ point, we find that the linear band touching is allowed. 
\begin{align}
[C_3]_p^{\dagger} H (\bm{k}) [C_3]_p = H (C_3^{-1}[\bm{k}]), ~~[R_x]_p^{\dagger} H(k_x, k_y) [R_x]_p = H(-k_x, k_y),
\end{align}
allows $H(\bm{k}) \propto k_x \hat{s}^x + k_y \hat{s}^y$ (where $(\hat{s}_x, \hat{s}_y)$ are the Pauli matrices obtained by properly rotating $\sigma^x$ and $\sigma^y$.). Hence, the quadratic band touching cannot be protected by $[C_3]_p$ and $[R_x]_p$. Nevertheless, within our network model, the touching is found to be robust though the touching is not protected by the symmetries. 

We can show that we need the six-fold rotation symmetry $[C_6]_p$ to protect the quadratic band touching and this is consistent with Ref. \cite{sun09}. Thus, on imposing $[C_6]_p$, we can fix the Hamiltonian as 
\begin{align}
H = \epsilon_0 (|\bm{k}|) + \Big( \frac{k_x^2 - k_y^2}{2m} \sigma^x + \frac{2k_x k_y}{2m} \sigma^y \Big).
\end{align}
To match the band spectrum seen in the model, we have $\epsilon_0 (|\bm{k}|) = \frac{\bm{k}^2}{2m}$ and thus the lower band is completely flat and the density of state at the zero energy is divergent.  

\subsection{Comparison with Twisted Graphene Bilayer}
Here we extend our discussion in the main text on the similarity between our network system and the theoretical models \cite{Efimkin, bist11} for the twisted graphene bilayers. In particular, we compare ours with the network model in Ref. \cite{Efimkin} and a continuum Dirac fermion model in Ref. \cite{bist11}.

To start with, we find that our network system is close to the network model appeared in Ref \cite{Efimkin}. In Ref. \cite{Efimkin}, the twisted graphene bilayer at a small twisting angle has been considered. When the twisting angle is small, there is a periodic array of domain walls separating the locally AA-stacked regions and the locally AB-stacked regions. Ref. \cite{Efimkin} argued that these domain walls trap localized one-dimensional metallic channels. These one-dimensional modes scatter at the nodes, which form a triangular lattice (in our case, the nodes form a honeycomb lattice). The structure of their model is quite similar to ours and indeed Ref. \cite{Efimkin} obtained a similar spectrum as ours: Dirac fermions, nearly flat bands, as well as van-Hove singularities.   

We can also make a comparison of our network system with the continuum Dirac theory of magic-angle twisted bilayer graphene in Ref. \cite{bist11}. In this approach, the Dirac fermions coming from the top and bottom layers interfere each other, and as a result, the bands become flat. Theoretically, this flat spectrum is speculated to be the source of surprising correlation-driven phenomena seen in the experiments \cite{cao18, lian18, po18}. One may note that our network model appears to be different than the continuum Dirac theory of Ref. \cite{bist11}. Despite of the difference in the theoretical treatments, we emphasize that our network model and the result of Ref. \cite{bist11} share the strikingly-similar features in spectrum: Dirac fermions, flat bands and associated singularities in density of states, which are believed to play an essential role in the correlation physics.

Both the twisted bilayer graphene and our honeycomb network have weak disorders \cite{brih12, raza16}. For example, there are some imperfect hexagons in our network and imperfect triangles in twisted bilayer graphene. Naively one expects that such weak disorders would immediately localize the flat bands and completely destroy associated many-body physics.  However, the previous study \cite{chal10} surprisingly found that the flat bands do not get immediately localized but become critical. This implies that the flat bands are stronger against disorders than we naively expect. Though a more thorough investigation is desirable, we expect from the reference \cite{chal10} that the flat bands retain relatively flat spectrum even with the weak disorders and hence is expected to remain very susceptible to many-body physics.

In summary, we have shown that the two systems, twisted graphene bilayer and our network system, share the surprising similarities including the flat bands and a large density of states, which are the key to the exotic correlation-driven phenomena.  

\subsection{Interlayer coupling}

In this subsection, we consider the effect of interlayer coupling to the electronic structures in the conducting network, and we will argue that, in general, the interlayer couplings between the layers will little affect the emergent electronic structures.

For the concrete-ness of our theoretical discussion, we first assume that the charge-density wave domains in the nearly commensurate phase remain insulating even after the inclusions of interlayer couplings [see SFig.5]. With this in hand, all the lowest-energy electronic states are in the domain walls in the conducting networks, and the interlayer couplings will introduce the coupling between these metallic modes inside the conducting networks living in different layers.

Among various possible couplings, the most important coupling, which can largely modify the band structure, is the electron hopping process between the layers. Note that this is proportional to the wavefunction overlap between the states of domain walls in different layers, and the states are highly localized within each domain walls. Hence, the effect of coupling will be strongly suppressed if not the networks are almost exactly overlapping to each other when seen from c-axis. From the available literature \cite{cho16}, we note that the networks in different layers are not correlated to each other and thus we expect that the emergent band structure of the low-energy theory will not be affected much by the interlayer coupling.

\newcommand{\AP}[3]{Adv.\ Phys.\ {\bf #1}, #2 (#3)}
\newcommand{\CMS}[3]{Comput.\ Mater.\ Sci. \ {\bf #1}, #2 (#3)}
\newcommand{\PR}[3]{Phys.\ Rev.\ {\bf #1}, #2 (#3)}
\newcommand{\PRL}[3]{Phys.\ Rev.\ Lett.\ {\bf #1}, #2 (#3)}
\newcommand{\PRB}[3]{Phys.\ Rev.\ B\ {\bf #1}, #2 (#3)}
\newcommand{\NA}[3]{Nature\ {\bf #1}, #2 (#3)}
\newcommand{\NAP}[3]{Nat.\ Phys.\ {\bf #1}, #2 (#3)}
\newcommand{\NAM}[3]{Nat.\ Mater.\ {\bf #1}, #2 (#3)}
\newcommand{\NAC}[3]{Nat.\ Commun.\ {\bf #1}, #2 (#3)}
\newcommand{\NAN}[3]{Nat.\ Nanotechnol.\ {\bf #1}, #2 (#3)}
\newcommand{\NARM}[3]{Nat.\ Rev.\ Mater.\ {\bf #1}, #2 (#3)}
\newcommand{\NT}[3]{Nanotechnology {\bf #1}, #2 (#3)}
\newcommand{\JP}[3]{J.\ Phys.\ {\bf #1}, #2 (#3)}
\newcommand{\JAP}[3]{J.\ Appl.\ Phys.\ {\bf #1}, #2 (#3)}
\newcommand{\JPSJ}[3]{J.\ Phys.\ Soc.\ Jpn.\ {\bf #1}, #2 (#3)}
\newcommand{\JPC}[3]{J.\ Phys.\ C: \ Solid State Phys.\ {\bf #1}, #2 (#3)}
\newcommand{\PNAS}[3]{Proc.\ Natl.\ Acad.\ Sci. {\bf #1}, #2 (#3)}
\newcommand{\PRSL}[3]{Proc.\ R.\ Soc.\ Lond. A {\bf #1}, #2 (#3)}
\newcommand{\PBC}[3]{Physica\ B+C\ {\bf #1}, #2 (#3)}
\newcommand{\PAC}[3]{Pure Appl.\ Chem. \ {\bf #1}, #2 (#3)}
\newcommand{\SCI}[3]{Science\ {\bf #1}, #2 (#3)}
\newcommand{\SCA}[3]{Sci.\  Adv.\ {\bf #1}, #2 (#3)}
\newcommand{\RPP}[3]{Rep.\ Prog.\ Phys. \ {\bf #1}, #2 (#3)}
\newcommand{\SCR}[3]{Sci.\  Ref.\ {\bf #1}, #2 (#3)}